\documentclass[aps,prd,floats,twocolumn,epsf,nofootinbib,showpacs,amssymb]{revtex4-1}
\pdfoutput=1
\usepackage{graphicx}
\usepackage{bm}
\usepackage{times}
\usepackage{slashed}
\usepackage{color}
\usepackage{slashed}
\usepackage{lipsum}
\usepackage{subfigure}
\usepackage{multirow}
\usepackage{amsmath}
\usepackage{array} 
\usepackage{varwidth} 

\newcommand{\be}{\begin{equation}}
\newcommand{\ee}{\end{equation}}
\newcommand{\bea}{\begin{eqnarray}}
\newcommand{\eea}{\end{eqnarray}}

\begin{document}

\title{Response to Comment on ``Dark Matter Annihilation Can Produce a Detectable Antihelium Flux through $\bar{\Lambda}_b$ Decays''}
\author{Martin Wolfgang Winkler}
\email{martin.winkler@su.se, ORCID: orcid.org/0000-0002-4436-0820}
\affiliation{Stockholm University and The Oskar Klein Centre for Cosmoparticle Physics,  Alba Nova, 10691 Stockholm, Sweden}
\author{Tim Linden}
\email{linden@fysik.su.se, ORCID: orcid.org/0000-0001-9888-0971}
\affiliation{Stockholm University and The Oskar Klein Centre for Cosmoparticle Physics,  Alba Nova, 10691 Stockholm, Sweden}

\begin{abstract}
\noindent
In a recent paper we showed that the decay of intermediate $\bar{\Lambda}_b$ baryons can dramatically enhance the antihelium flux from dark matter annihilation. Our antihelium predictions were derived using several implementations of the Pythia and Herwig event generators which were calibrated to existing data on antideuteron and antihelium formation. Kachelrie{\ss} et al. have argued for a smaller antihelium flux compared to our most optimistic Monte Carlo model. However, we show that the arguments by Kachelrie{\ss} et al. are either incorrect or irrelevant for antihelium formation. Thus, the results of our original paper remain unchanged.
\end{abstract}

\maketitle

\begin{table*}[t]
\begin{center}
\begin{tabular}{|ccccc|}
\hline
  experiment & channel & measurement & Pythia (default) & Pythia ($\Lambda_b$-tune)\\[0.5mm]
\hline 
 &&&&\\[-3mm]
 LEP~\cite{Caso:1998tx,Abbiendi:1998ip} & $f(b\rightarrow \Lambda_b)$ & $0.101^{+0.039}_{-0.031}$ & 0.037 & 0.101 \\[1.5mm]
 LEP~\cite{Abbaneo:2001bv} & $f(b\rightarrow \Lambda_b,\Xi_b,\,\Omega_b)$ & $0.117\pm 0.021$ & 0.047 & 0.127 \\[1.5mm]
 Tevatron CDF~\cite{Aaltonen:2008zd} & $\frac{f(b\rightarrow \Lambda_b)}{f(b\rightarrow B)}$ & $0.281^{+0.141}_{-0.103}$ & 0.046 & 0.135 \\[1.5mm]
 LHCb~\cite{Aaij:2019pqz} & $\frac{f(b\rightarrow \Lambda_b)}{f(b\rightarrow B)}$ & $0.259\pm 0.018$ & 0.048 & 0.134 \\[0.5mm]
 \hline
\end{tabular}
\end{center}
\caption{Measurements of $\Lambda_b$-production in various experiments compared to the prediction in default Pythia and the Pythia $\Lambda_b$-tune. The Pythia predictions have been adjusted to the specific kinematical ranges employed in the measurements.}
\label{tab:fbaryon}
\end{table*}

\noindent
\emph{Summary of Original Paper} -- The AMS-02 cosmic-ray experiment has tentatively detected a handful of cosmic-ray antihelium events~\cite{AMSLaPalma}. This observation is puzzling, since neither astrophysical processes nor dark matter annihilation were thought to produce a detectable antihelium flux. In recent work~\cite{Winkler:2020ltd}, we 
demonstrated that a previously neglected standard-model process -- the production of antihelium nuclei through the decay of intermediate $\bar{\Lambda}_b$ baryons -- could 
dramatically enhance the antihelium production rate from dark matter annihilation. 
The key insight is that $\bar{\Lambda}_b$, due to its antibaryon number and 5.6~GeV rest-mass, efficiently decays to multi-antibaryons states with small relative momentum which coalesce into antihelium.

The antihelium flux from $\bar{\Lambda}_b$ decay depends on (1) the rate of $\bar{\Lambda}_b$ 
baryon production in dark matter annihilations to bottom quarks, and (2) the probability of a single  $\bar{\Lambda}_b$ decaying into an antihelium nucleus. 
Since (2) has not been measured, our antihelium predictions are based on four different configurations of the Pythia and Herwig event generators. These models reflect the major uncertainties in antihelium formation due to the underlying hadronization models. Despite large numerical differences, three Monte Carlo implementations support a drastic enhancement of the antihelium yield through $\bar{\Lambda}_b$ decays by one/two orders of magnitude.\\

\noindent\emph{The Comment of Kachelrie{\ss} et al.} --
Recently, Kachelrie{\ss} et al. (hereafter KOT21)~\cite{Kachelriess:2021vrh}, raised two concerns regarding these results. The first targets the Pythia $\Lambda_b$-tune -- the event generator configuration that provides our most optimistic antihelium prediction. This tune increases the diquark formation parameter to {\tt probQQtoQ}=0.24 (Pythia default: 0.09) in order to rectify the underproduction of $\Lambda_{b}$ baryons in Pythia's default configuration. KOT21 makes two claims, arguing that the $\Lambda_b$-tune model:
\begin{itemize}
    \item[(1)] is not motivated because the default Pythia $b\rightarrow\Lambda_b$ transition rate is consistent with data (quoting agreement with LEP to within 1$\sigma$).
    \item[(2)] overpredicts light baryon formation (quoting a discrepancy of 33$\sigma$) and therefore overestimates antihelium formation.
\end{itemize}

As we will show below, these statements are either incorrect or inapplicable to our study. Moreover, we stress that even if the analysis of KOT21 were entirely correct, these criticisms amount to only a factor of $\sim$3~adjustment in a novel factor of $\sim$100~effect first pointed out in our original paper. 

First, statement (1) is incorrect. The fraction \mbox{$f(b\rightarrow \Lambda_b)$} has been determined by a variety of LEP~\cite{Caso:1998tx,Abbiendi:1998ip,Abbaneo:2001bv}, Tevatron~\cite{Aaltonen:2008zd} and LHC~\cite{Aaij:2019pqz} measurements. Several of the most relevant results are listed in Table~\ref{tab:fbaryon}, together with the predicted rate from both default Pythia models and our \mbox{$\Lambda_b$-tune} model. Standard Pythia implementations predict a $\Lambda_b$-production rate that falls below results from several independent experiments and channels at a combined significance exceeding $10\sigma$. Our $\Lambda_b$-tune model, on the other hand, provides a good fit to LEP data. It still underpredicts the data from hadron colliders, signaling that -- if anything -- the antihelium production rate in the $\Lambda_b$-tune model is on the conservative side.

Statement (2) is partially correct, but not relevant. While our $\Lambda_b$-tune model clearly rectifies heavy baryon production in Pythia, it does overestimate light baryon spectra. We first note that such a choice is well-motivated as antihelium production strongly depends on heavy baryonic processes. 

Additionally, and in contrast to claim (2), the offset in the light baryon spectra does not propagate to an increase in antihelium prediction. This is because all Monte Carlo configurations we employed have been fit to experimental data on antideuteron and antihelium formation. For the $\Lambda_b$-tune model, a significantly smaller coalescence momentum has been derived compared to default Pythia implementations. This effectively renormalizes light (anti)baryon spectra, compensating for their overproduction. This becomes obvious if one compares the decay rate $\bar{\Lambda}_b\rightarrow \overline{\text{He}}$ which agrees between default Pythia and the $\Lambda_b$-tune within an $\mathcal{O}(1)$-factor.\\

\noindent
The second concern raised by KOT21 is that Pythia overestimates the branching ratio $\bar{\Lambda}_b \rightarrow \overline{\text{He}}$. In support of this hypothesis KOT21 quote a $42\sigma$ mismatch (corresponding to a factor of $\sim 6$) between Pythia's prediction and the actual measurement of the process $\bar{\Lambda}_b \rightarrow \bar{\Lambda}_c^-+ \bar{p}p\pi^+$. We first stress that this process does not in any way contribute to antihelium formation. However, it does emerge through the extraction of a diquark-antidiquark pair from the vacuum in the hadronization stage -- a feature it shares with $\bar{\Lambda}_b \rightarrow \overline{\text{He}}$. KOT21 argue that the offset in $\bar{\Lambda}_b \rightarrow \bar{\Lambda}_c^-+ \bar{p}p\pi^+$ signals a bias in Pythia towards diquark formation in $\bar{\Lambda}_b$ decays, which could then also hint at an overestimate of the antihelium yield.

However, KOT21 fail to appreciate that an offset in a single decay rate could point to mismodeling in any number of relevant routines, some of which relate to diquark formation, and others which do not. In order to test whether the offset in $\bar{\Lambda}_b \rightarrow \bar{\Lambda}_c^-+ \bar{p}p\pi^+$  is linked to the mismodeling of  diquark formation, it is imperative to examine complementary processes that do not include diquark formation. 

In Table~\ref{tab:branchings} we show that Pythia, in fact, produces very similar offsets (factor of $\sim$6) in the rates $\bar{\Lambda}_b \rightarrow \bar{\Lambda}_c^-+ \pi^- \pi^+ \pi^+$ and $\bar{\Lambda}_b \rightarrow \bar{\Lambda}_c^-+ K^- K^+\pi^+$. Similarly to the process examined by KOT21 ($\bar{\Lambda}_b \rightarrow \bar{\Lambda}_c^-+ \bar{p}p\pi^+$) these processes include $\Lambda_b\rightarrow \Lambda_c$. However, they do not involve diquark formation (no baryon-antibaryon pair is produced). Thus, the similarity of these offsets hints at a mismodeling of the $\Lambda_b\rightarrow \Lambda_c$ transition in Pythia. While further analysis would be necessary to prove this hypothesis, it is already clear that -- in contrast to the claim of KOT21 -- the study of $\bar{\Lambda}_b \rightarrow \bar{\Lambda}_c^-+ \bar{p}p\pi^+$ can not be directly employed to draw conclusions regarding the accuracy of antihelium formation in Pythia. \\

\begin{table}[htp]
\begin{center}
\begin{tabular}{|ccc|}
\hline
   branching ratio & measurement & Pythia\\[0.5mm]
\hline 
 &&\\[-3mm]
  $\bar{\Lambda}_b \rightarrow \bar{\Lambda}_c^-+ \bar{p}\,p\,\pi^+$ & $(2.65\pm 0.29)\times 10^{-4}$ & $1.5 \times 10^{-3}$ \\[1.5mm]
  $\bar{\Lambda}_b \rightarrow \bar{\Lambda}_c^-+ \pi^- \pi^+ \pi^+$ & $(7.7\pm 1.1)\times 10^{-3}$ & $5.1 \times 10^{-2}$ \\[1.5mm]
  $\bar{\Lambda}_b \rightarrow \bar{\Lambda}_c^-+ K^- K^+\pi^+$ &  $(1.02\pm 0.12)\times 10^{-3}$ & $4.4 \times 10^{-3}$ \\[1.5mm]
 \hline
\end{tabular}
\end{center}
\caption{Measured branching ratios of $\bar{\Lambda}_b$ from~\cite{Aaij:2018bre,Aaij:2020zzw,Zyla:2020zbs} compared to the Pythia prediction.}
\label{tab:branchings}
\end{table}

\noindent\emph{Concluding Remarks} -- While KOT21 make two criticisms regarding the usage of Pythia models within our work, their arguments do not actually target any of the main conclusions of our paper. In particular, they neither challenge the novel antihelium mechanism that we examine, nor its importance for $\overline{\text{He}}$-formation. Rather KOT21 argues for a smaller antihelium flux compared to the most optimistic estimate from our original paper. The main concern applies to one particular Monte Carlo implementation, the Pythia $\Lambda_b$-tune. While this model predicts the highest antihelium yield, an independent Herwig implementation (not examined by KOT21) only falls short by a factor of 3. Hence, KOT21 boils down to a discussion of an $\mathcal{O}(1)$ factor in a novel factor of $\sim$100 effect.

Aside from their minor importance, the criticisms in KOT21 are based on offsets between Pythia implementations and measured decay rates in certain channels. We have shown that these offsets either (i) concern decay rates that are irrelevant to antihelium formation, or (ii) have explicitly been accounted for in our work. Therefore -- contrary to the claim of the authors -- none of the arguments provided by KOT21 suggests any reduction of the antihelium yield.

As a final remark, we agree that event generators cannot replace an actual measurement of the transition $\bar{\Lambda}_b\rightarrow \overline{\text{He}}$ -- a measurement that we hope to stimulate by our simulation work. However, our original work contains a balanced discussion of the underlying uncertainties in the antihelium predictions. In particular, we show results from two different event generators, explore a large parameter space of potential input parameters and modeling decisions, and show the resulting antihelium flux in each model. The culmination of this evidence supports our original claim that $\bar{\Lambda}_b$ decays may significantly enhance the antihelium formation rate in dark matter annihilation events -- an exciting possibility given recent AMS-02 claims of a detectable antihelium flux.

\section{Acknowledgements}
We thank John Beacom, Dan Hooper, Michael Kachelrie{\ss}, Sergey Ostapchenko and Jonas Tjemsland for helpful conversations. TL is partially supported by the Swedish Research Council under contract 2019-05135, the Swedish National Space Agency under contract 117/19 and the European Research Council under grant 742104. 


%

\end{document}